\documentclass[runningheads]{llncs}
\usepackage{graphicx}

\usepackage{tikz}
\usepackage{comment}
\usepackage{amsmath,amssymb} 
\usepackage{amsbsy}
\usepackage{color}

\usepackage[accsupp]{axessibility}  

\usepackage{array,multirow}
\newcolumntype{P}[1]{>{\centering\arraybackslash}p{#1}}
\newcolumntype{M}[1]{>{\centering\arraybackslash}m{#1}}
\usepackage{mathrsfs} 
\usepackage{algorithm}
\usepackage{algpseudocode}
\usepackage{cite}
\usepackage{booktabs}
\usepackage{wrapfig}
\usepackage{caption}
\usepackage{orcidlink}

\begin{document}
	
	\pagestyle{headings}
	\mainmatter
	\def\ECCVSubNumber{6455}  
	
	\title{DiffuseMorph: Unsupervised Deformable Image Registration Using Diffusion Model} 
	
	
	\titlerunning{Unsupervised Deformable Image Registration Using Diffusion Model}
	%
	\author{Boah Kim\orcidlink{0000-0001-6178-9357}\index{Kim, Boah} \and 
		Inhwa Han\orcidlink{0000-0002-8666-8313}\index{Han, Inhwa} \and
		Jong Chul Ye\orcidlink{0000-0001-9763-9609}\index{Ye, Jong Chul}} 
	\authorrunning{B. Kim et al.}
	\institute{Korea Advanced Institute of Science and Technology (KAIST), Daejeon, South Korea \\
		\email{\{boahkim, inhwahan, jong.ye\}@kaist.ac.kr}}
	
	\maketitle
	
	\begin{abstract}
		Deformable image registration is one of the fundamental tasks in medical imaging. Classical registration algorithms usually require a high computational cost for iterative optimizations. Although deep-learning-based methods have been developed for fast image registration, it is still challenging to obtain realistic continuous deformations from a moving image to a fixed image with less topological folding problem. To address this, here we present a novel diffusion-model-based image registration method, called DiffuseMorph. DiffuseMorph not only generates synthetic deformed images through reverse diffusion but also allows image registration by deformation fields. Specifically, the deformation fields are generated by the conditional score function of the deformation between the moving and fixed images, so that the registration can be performed from continuous deformation by simply scaling the latent feature of the score. Experimental results on 2D facial and 3D medical image registration tasks demonstrate that our method provides flexible deformations with topology preservation capability.
		
		\keywords{Image registration \and diffusion model \and image deformation \and unsupervised learning}
	\end{abstract}

	\section{Introduction}
	\label{sec:intro}
		\let\thefootnote\relax\footnotetext{
		Part of this paper is presented at the 25th International Conference on Medical Image Computing and Computer Assisted Intervention, MICCAI 2022 \cite{kim2022diffusion}.
	}
	Deformable image registration is to estimate non-rigid voxel correspondences between moving and fixed image pairs. This is especially important for medical image analysis such as disease diagnosis and treatment monitoring, since the anatomical structures or shapes of medical images are different according to subjects, scanning time, imaging modality, etc. Accordingly, various image registration methods have been studied over the past decades.
	
	Classical image registration approaches usually attempt to align images by solving a computationally expensive optimization problem \cite{avants2008symmetric, ashburner2007fast, klein2009elastix}. To address this computational issue, deep-learning-based image registration methods have been extensively studied \cite{cao2017deformable, balakrishnan2018unsupervised, onofrey2013semi, lei20204d, kim2021cyclemorph}, which train neural networks to estimate the registration field by taking the moving and fixed images as network inputs. These approaches provide fast deformation while maintaining registration accuracy. However, the supervised methods usually require the ground-truth registration fields \cite{onofrey2013semi, rohe2017svf}, and some of the existing unsupervised approaches need additional diffeomorphic constraints \cite{dalca2018unsupervised, krebs2018unsupervised} or the cycle-consistency \cite{kim2021cyclemorph} for topology preservation.
	
	\begin{figure}[t!]
		\centering
		\includegraphics[width=\linewidth]{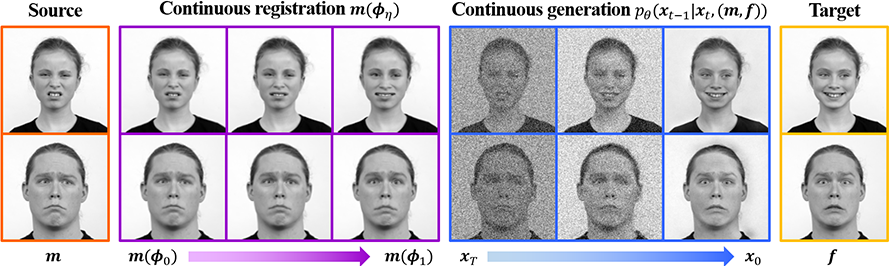}
		\caption{DiffuseMorph provides not only deformable image registration along the continuous trajectory by simply scaling the latent features in generating deformation fields but also synthetic deformed images through continuous generation by the reverse diffusion process. }
		\label{fig:representative}
	\end{figure}
	
	Recently, score-based diffusion models have shown high-quality performance in image generation \cite{song2020score, vahdat2021score}. In particular, the denoising diffusion probabilistic model (DDPM) \cite{ho2020denoising, sohl2015deep} learns the Markov transformation from Gaussian noise to data distribution and provides diverse samples through the stochastic diffusion process by estimating the latent feature of score function, which has been applied to many areas of computer vision \cite{song2020denoising, fadnavis2020patch2self, chen2020wavegrad, jeong2021diff, sasaki2021unit}. To generate images with desired semantics, conditional denoising diffusion models have been also presented \cite{saharia2021image, choi2021ilvr}. However, it is challenging to apply DDPM to the image registration task, since the proper registration should be performed through the deformation field for the moving image rather than image generation.
	
	In this paper, by leveraging the property of the diffusion model where the estimated latent feature provides spatial information to generate images, we present a novel unsupervised deformable image registration approach, dubbed DiffuseMorph, by adapting the DDPM to generate deformation fields. Specifically, our proposed model is composed of a diffusion network and a deformation network: the former network learns a conditional score function of the deformation between moving and fixed images, and the latter network estimates the deformation field using the latent feature from the score function and provides deformed images. These two networks are jointly trained in an end-to-end learning manner so that DiffuseMorph not only estimates Markov transformation in the direction in which the moving image is deformed into the fixed image, but also produces the registration field for the moving image to be warped into the fixed image. Since the latent feature from the conditional score function of the  diffusion model has spatial information of the condition, the linear scaling of the latent feature may provide deformation fields along the continuous trajectory from the moving to the fixed images.
	
	Accordingly, as shown in Fig.~\ref{fig:representative}, the proposed DiffuseMorph allows both image registration along the continuous trajectory and synthetic deformed image generation. Specifically, our trained model provide the continuous deformations from the moving image to the fixed image by simply interpolating the latent feature that is used as an input for the deformation network. In addition, the proposed model can quickly generate synthetic deformed images similar to the fixed images. Here, to further accelerate the diffusion procedure, instead of starting from random Gaussian noise, we present a generative process in which the moving image is propagated one step via forward diffusion and then iteratively refined through the reverse diffusion process of the DDPM. This reduces the number of diffusion steps significantly and makes the sample retain the original moving image content.
	
	We demonstrate the performance of the proposed method on 2D facial expression registration and 3D medical image registration tasks. The experimental results verify that our model achieves high performance in registration accuracy. Also, thanks to the latent feature estimated from the diffusion model, our method enables real-time image registration along the continuous trajectory between the moving and fixed images, which is more realistic than the comparative learning-based registration methods. Our main contributions are summarized as:
	\begin{itemize}
		\item We propose DiffuseMorph, the first image registration method employing the denoising diffusion model conditioned on a pair of moving and fixed images.
		\item When the proposed model is trained, our model not only performs image registration along the continuous trajectory from the moving to fixed images by scaling the latent feature but also generates synthetic deformed images through the fast reverse diffusion process.
		\item We demonstrate that the proposed method can be applied to 2D and 3D image registration tasks and provide accurate deformation with comparable topology preservation over the existing methods.
	\end{itemize}

	\section{Backgrounds and Related Works}
	\label{sec:background}
	
	\subsection{Deformable Image Registration}
	
	Given a moving image $m$ and a fixed image $f$, classical deformable image registration methods are performed by solving the following optimization problem:
	\begin{align}
		\label{eq:origdeformloss}
		\phi^\ast = \underset{\phi}{\mathrm{argmin}} \quad  L_{sim} (m(\phi), f)+L_{reg} (\phi),
	\end{align}
	where $\phi^\ast$ is the optimal registration field to deform the moving image into the fixed image. $L_{sim}$ is the dissimilarity function to compute the similarity between the deformed and fixed images, and $L_{reg}$ is the regularization penalty of the registration field. By minimizing the energy function, the deformed image $m(\phi)$ is estimated by warping the moving image. In particular, diffeomorphic registration can be achieved when one imposes additional constraints on the field $\phi$ such that the deformation mapping is differentiable and invertible, thereby preserving the topology \cite{beg2005computing, vercauteren2009diffeomorphic, avants2008symmetric}.
	
	\subsubsection{Learning-based Registration Methods}
	As the traditional registration approaches usually require large computation and long runtime, deep learning methods have been extensively studied lately, which estimates the deformation field in real time once a neural network is trained. However, supervised learning methods train the networks using the ground-truth registration fields \cite{cao2017deformable, yang2017quicksilver, rohe2017svf, cao2018deep}, which needs high-quality labels for training. To alleviate this, weakly-supervised registration models that use pseudo-labels such as segmentation maps have been developed \cite{hu2018weakly, xu2019deepatlas}. On the other hand, unsupervised learning approaches train the networks by computing similarity between the deformed image and the fixed reference \cite{balakrishnan2018unsupervised, mahapatra2018deformable, de2019deep, lei20204d, kim2021cyclemorph}. To guarantee topology preservation, learning-based diffeomorphic registration methods are also presented \cite{dalca2018unsupervised, krebs2018unsupervised, dalca2019unsupervised}, which have the layer of scaling and squaring integration for the diffeormorphic constraint.
	
	These existing methods may provide intermediate deformations between the moving and fixed images by scaling the registration field or integrating the velocity field in shorter timescales. However, our method produces more realistic continuous deformation by scaling the latent feature that has spatial information of the moving and fixed images, improving the performance of image registration.

	\subsection{Denoising Diffusion Probabilistic Model}
	Recently, the denoising diffusion probabilistic model (DDPM) \cite{ho2020denoising, sohl2015deep}, one of the generative models, is presented to learn the Markov transformation from the simple Gaussian distribution to the data distribution. In the forward diffusion process, noises are gradually added to the data $x_0$ using a Markov chain, in which each step of sampling latent variables $x_t$ for $t\in [0, T]$ is defined as a Gaussian transition:
	\begin{align}
		q(x_t | x_{t-1} )=\mathcal{N}(x_t;\sqrt{1-\beta_t} x_{t-1},\beta_t I),
	\end{align}
	where $0<\beta_t<1$ is a variance of the noise. The resulting distribution of $x_t$ given $x_0$ is then expressed as:
	\begin{align}
		q(x_t | x_0 )=\mathcal{N}(x_t;\sqrt{\alpha_t} x_0,(1-\alpha_t)I),
	\end{align}
	where $\alpha_t=\Pi_{s=1}^t(1-\beta_s)$. Accordingly, given $\epsilon\sim\mathcal{N}(0, I)$, $x_t$ can be sampled by:
	\begin{align}
		x_t = \sqrt{\alpha_t}x_0 + \sqrt{1-\alpha_t} {\epsilon}.
		\label{eq:ddpm_xt}
	\end{align}
	
	For the generative process to perform the reverse diffusion, DDPM learns the parameterized Gaussian process $p_\theta (x_{t-1} | x_t)$, which is represented as:
	\begin{align}
		p_\theta (x_{t-1} | x_t)=\mathcal{N}(x_{t-1};\mu_\theta (x_t,t),\sigma_t^2 I),
	\end{align}
	where $\sigma_t$ is a fixed variance, and $\mu_\theta (x_t,t)$ is a learned mean defined as:
	\begin{align}\label{eq:condmean}
		\mu_\theta (x_t,t) = \frac{1}{\sqrt{1-\beta_t}} \left(x_t-\frac{\beta_t}{\sqrt{1-\alpha_t}} \epsilon_\theta (x_t,t)\right),
	\end{align}
	where $\epsilon_\theta$ is a parameterized model. In fact, the model $\epsilon_\theta (x_t,t)$ is just a scaled version of the score function $s_\theta(x_t,t)$ \cite{song2019generative}, which is the gradient of the $\log p_\theta (x_t)$.
	Once the model $\epsilon_\theta$ is trained, the data is sampled by the following stochastic generation step:
	$x_{t-1}=\mu_\theta (x_t,t) +\sigma_t z$, where $\sigma_t^2=\frac{1-\alpha_{t-1}}{1-\alpha_t}\beta_t$ and $z\sim\mathcal{N}(0, I)$.

	\subsubsection{Conditional Diffusion Models} 
	In order to generate images with desired semantics, conditional diffusion models have been proposed recently \cite{song2020score, song2020denoising, saharia2021image, choi2021ilvr, ho2022cascaded, nichol2021improved}, which gives the conditional reference image to the network or the generative process. DDIM \cite{song2020denoising} proposes a deterministic non-Markovian generative process starting from an initial condition to control the image generation of the reverse diffusion process. SR3 \cite{saharia2021image} presents a method to train the DDPM with a conditioned image for the super-resolution task. ILVR \cite{choi2021ilvr} proposes a conditioning iterative generative process using an unconditional model. However, these diffusion model-based generative methods are concerned about image generation, and cannot be used for image registration as they do not produce any deformation field for the registration.

	\begin{figure}[t!]
		\centering
		\includegraphics[width=\linewidth]{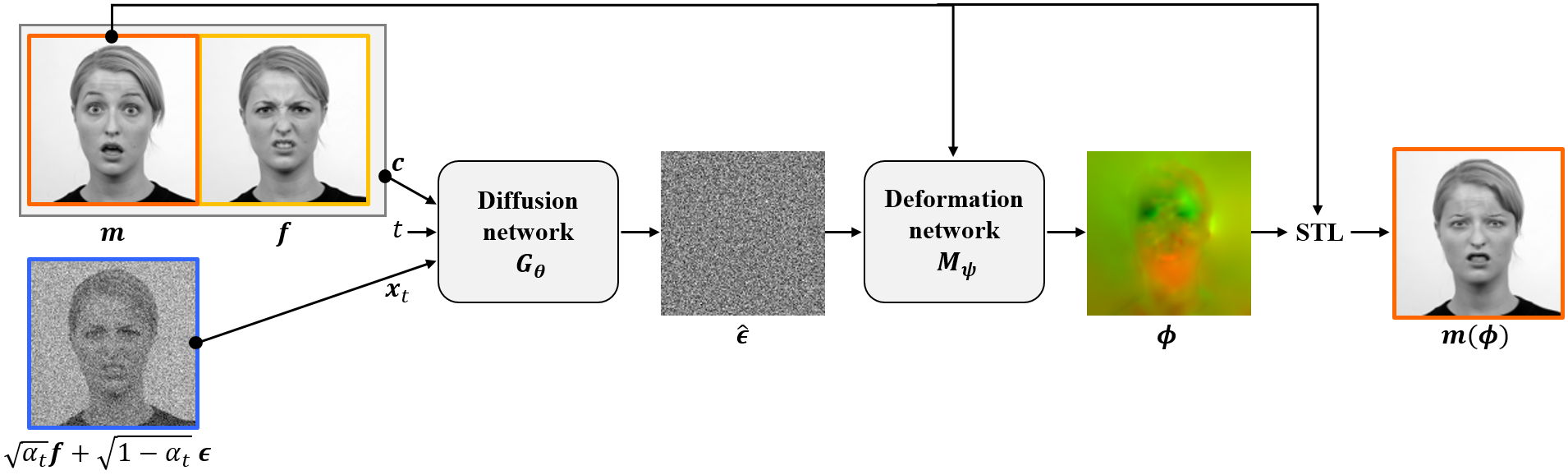}
		\caption{The training framework of DiffuseMorph. Given a condition with a pair of a moving image $m$ and a fixed image $f$, the diffusion network $G_\theta$ estimates the conditional score function of the deformation, and the deformation network $M_\psi$ outputs the registration field $\phi$. Then, using the spatial transformation layer (STL), the moving image is warped into the fixed image. }
		\label{fig:method_train}
	\end{figure}
	
	\section{Proposed Method}
	\label{sec:method}
	
	\subsection{Framework of DiffuseMorph}
	By leveraging the capability of DDPM, we aim to develop a novel diffusion model-based unsupervised image registration approach. Since the image registration is to warp the moving image using the deformation field, we design our model with two networks as illustrated in Fig.~\ref{fig:method_train}: one is a diffusion network $G_\theta$ to estimate a conditional score function, and the other is a deformation network $M_\psi$ that actually outputs the registration field using the score function.
	
	Specifically, for the moving source image $m$ and the fixed reference image $f$, the diffusion network $G_\theta$ is trained to learn the conditional score function of the  deformation between the moving and fixed images given the condition $c=(m, f)$. For this, we sample the latent variable $x_t$ of the target by (\ref{eq:ddpm_xt}), defining the fixed image as the target, i.e. $x_0=f$. Moreover, to make the network $G_\theta$ aware of the level of noise, we also give the number of time steps for the noise to the network, similar to \cite{ho2020denoising}. 
	
	On the other hand, the deformation network $M_\psi$ takes the latent feature of the conditional score function $\hat{\epsilon}$ that is an output of the diffusion network, as well as the moving source image $m$. Then the network outputs the registration field $\phi$, providing the deformed image $m(\phi)$ by warping the moving image $m$ using the spatial transformation layer (STL) \cite{jaderberg2015spatial}. To deform 2D/3D images in our experiments, we adopt the transformation function using bi-/tri-linear interpolation. 
	
	\subsection{Loss Function}
	Recall that the diffusion network $G_\theta$ and the deformation network $M_\psi$ are jointly trained in an end-to-end learning manner. Thus, for the training of our model, we design the objective function as follows:
	\begin{align} 
		\label{eq:loss_tot}
		\min_{G_\theta,M_\psi} L_{diffusion}(c,x_t,t)+ \lambda L_{regist}(m,f),
	\end{align}
	where $L_{diffusion}$ and $L_{regist}$ are the diffusion loss and the registration loss, respectively, and $\lambda$ is a hyper-parameter. The detailed description of each loss function is as follows.
	
	Given the condition $c$ and the perturbed data $x_t$ at the time step $t \in [0, T_{train}]$, the diffusion loss is to learn the conditional score function:
	\begin{align}
		L_{diffusion} (c,x_t,t)=\mathbb{E}_{\epsilon,x_t,t}||G_\theta (c, x_t, t)-\epsilon||_2^2,
	\end{align}
	where $\epsilon\sim \mathcal{N}(0, I)$. Also, the registration loss is to estimate the deformation field so that the deformed source image has similar shape of the fixed image, which is designed as the traditional energy function in (\ref{eq:origdeformloss}):
	\begin{align}
		\label{eq:loss_regist}
		L_{regist} (m,f) = -(m(\phi) \otimes f) + \lambda_\phi \sum ||\nabla \phi ||^2,
	\end{align}
	where $\phi=M_\psi (m, \hat\epsilon)$ with $\hat\epsilon$ referring to the diffusion network output, and $\lambda_\phi$ is a hyper-parameter. The first term of (\ref{eq:loss_regist}) is the local normalized cross-correlation \cite{balakrishnan2018unsupervised} between the deformed image and fixed image, and the second term is the smoothness penalty on the registration field. We set $\lambda_\phi=1$.
	
	It is remarkable that the net effect of the two loss functions is that $G_\theta$ is trained to estimate the latent feature for the conditional score function of the deformation, which has the spatial information of the moving and fixed images. Accordingly, the latent feature helps the proposed model to perform image registration along the continuous trajectory, which can provide topology preservation. Furthermore, when combined with the reverse diffusion process, the latent feature guides the reverse diffusion to generate the synthetic deformed image from the moving image initialization.

	\begin{figure}[t!]
		\centering
		\includegraphics[width=\linewidth]{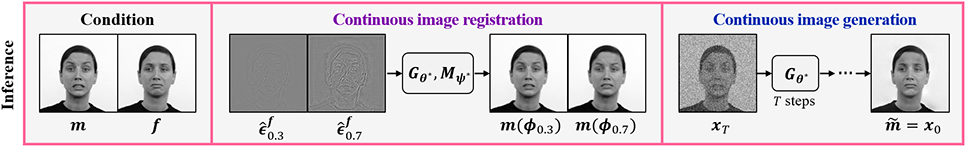}
		\caption{In the inference phase, our model provides not only the image registration $m(\phi)$ that warps the moving image, but also generates synethetic
			images $\tilde{m}$.}
		\label{fig:method_infer}
	\end{figure}
	
	\subsection{Image Registration Using DiffuseMorph} 
	When the networks of the proposed model are trained, in the inference phase, they provide image registration by estimating the deformation field for the moving image to be aligned with the fixed image. Thanks to the end-to-end training of our model, the diffusion network allows the deformation network to generate the regular registration field. Specifically, using the learned parameters of $G_{\theta^\ast}$ and $M_{\psi^\ast}$, the registration field $\phi$ at $t=0$ is estimated by:
	\begin{align}
		\label{eq:infer_def}
		\phi = M_{\psi^\ast} (m, G_{\theta^\ast} (c,x_0, t)),
	\end{align}
	where $x_0$ is set to the fixed target image $f$. Then, the deformed image $m(\phi)$ is computed using the estimated field $\phi$ through the spatial transformation layer. Therefore, our model performs image registration at a single step with the smooth registration field.
	
	\subsubsection{Image Registration Along Continuous Trajectory}
	
	In the image registration that warps the moving image into the fixed image, our model provides the
	continuous deformation of the moving image along the trajectory toward the fixed image. This is possible since the deformation network estimates the registration field according to the latent feature. Specifically, if the latent feature from the conditional score is set to zero, the deformation network outputs the registration field that hardly warps the moving image, whereas when the latent feature is given as in (\ref{eq:infer_def}), the  deformation network estimates the registration field that deforms the moving image into the fixed image.
	
	\begin{wrapfigure}[11]{r}{0.57\textwidth}
		\begin{minipage}[t][2cm][c]{0.57\textwidth}
			\begin{algorithm}[H]
				\caption{Continuous image registration}
				\label{alg:inference1}
				\begin{algorithmic}[1]
					\State \textbf{Input}: Conditional images, $c=(m, f)$
					\State \textbf{Output}: Deformed moving image, $m(\phi_\eta)$ 
					\State Set the latent feature $\hat{\epsilon}^f=G_{\theta^\ast}(c, f, 0)$
					\For{$\eta \in [0, 1]$}
					\State $\hat{\epsilon}_\eta^f \leftarrow \eta \cdot \hat{\epsilon}^f$
					\State $\phi_\eta \leftarrow M_{\psi^\ast} (m,  \hat{\epsilon}_\eta^f)$
					\EndFor  =\\
					\Return $m(\phi_\eta)$
				\end{algorithmic}
			\end{algorithm}
		\end{minipage}
	\end{wrapfigure}
	
	Accordingly, as described in Algorithm~\ref{alg:inference1}, for the latent feature $\hat{\epsilon}^f=G_{\theta^\ast} (c, f, 0)$, the registration field $\phi_\eta$ for the continuous image deformation can be generated by simply interpolating the latent feature:
	\begin{align}
		\label{eq:infer_def2}
		\phi_\eta = M_{\psi^\ast} (m,  \hat{\epsilon}_\eta^f), 
	\end{align}
	where $\hat{\epsilon}_\eta^f = \eta \cdot \hat{\epsilon}^f$ for $0\leq \eta \leq 1$. 
	We believe that this interesting phenomenon occurs from learning the conditional score function of deformation, as will be observed later in our experiments.

	\subsubsection{Synthetic Image Generation via Reverse Diffusion}
\begin{figure}[t!]
		\begin{minipage}[t][4cm][c]{0.35\linewidth}
			\centering
			\includegraphics[width=\linewidth]{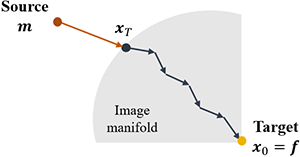}
			\caption{Generative process toward fixed target data distribution.}
			\label{fig:method_infer_gen}
		\end{minipage}\hfill
		\begin{minipage}[t][3cm][c]{0.63\linewidth}
			\begin{algorithm}[H]
				\caption{Synthetic image generation process}
				\label{alg:inference}
				\begin{algorithmic}[1]
					\State \textbf{Input}: Conditional images, $c=(m, f)$
					\State \textbf{Output}: Synthetic deformed image, $x$ 
					\State Set $T \in (0, T_{train})$
					\State Sample ${x}_T\!=\!\sqrt{\alpha_T} m\! +\! \sqrt{1-\alpha_T}\epsilon$,\! where $\epsilon\! \sim \!\mathcal{N}(0, I)$
					\For{$t=T, T-1, ..., 1$}
					\State $z \sim \mathcal{N}(0, I)$
					\State $x_{t-1}\leftarrow\frac{1}{\sqrt{1-\beta_t}} (x_t-\frac{\beta_t}{\sqrt{1-\alpha_t}} G_{\theta^\ast} (c, x_t, t))+\sigma_t z$
					\EndFor \\
					\Return $x_0$
				\end{algorithmic}
			\end{algorithm}
		\end{minipage}
	\end{figure}

	In our method, the latent feature generated by the diffusion network itself also guides the generation of synthetic deformed images through the reverse diffusion process. Here, as the diffusion network learns the conditional score function for the deformation between the moving image and the fixed image, our image generation starts from the moving image, in contrast to the existing conditional generative process of DDPM \cite{choi2021ilvr, saharia2021image} that starts from the pure Gaussian noise $x_T\sim \mathcal{N}(0,I)$. When we set the initial state with the original moving image $m$, the one-step forward diffusion is performed by:
	\begin{align}
		x_T = \sqrt{\alpha_T}m + \sqrt{1-\alpha_T}\epsilon,
	\end{align}
	where $\epsilon \sim \mathcal{N}(0, I)$, and $\alpha_T$ is the noise level at the time step $T\leq T_{train}$. Here, the time step $T$ is set to a value not to lose the identity of the image. This forward sampling allows the moving image distribution to be close to the fixed image distribution, as illustrated in Fig.~\ref{fig:method_infer_gen}, which reduces the number of reverse diffusion steps and generation time.
	
	Then, by starting from $x_T$, the generation of the synthetic image $x_0$ that fits into the fixed image $f$ is performed by the following reverse diffusion process from $t=T$ to $t=1$:
	\begin{align}
		x_{t-1}=\frac{1}{\sqrt{1-\beta_t}} \left(x_t-\frac{\beta_t}{\sqrt{1-\alpha_t}} G_{\theta^\ast}(c,x_t,t)\right)+\sigma_t z,
	\end{align}
	where $z \sim \mathcal{N}(0, I)$. Here, in choosing the total steps of reverse diffusion, we employ \cite{chen2020wavegrad} that presents an efficient inference method. Thus, one can flexibly set the number of sampling steps, and in our experiments, we set the reverse steps as 200 in maximum. The pseudocode of this generative process of DiffuseMorph is described in Algorithm~\ref{alg:inference}. 
	
	\section{Experimental Results}
	\label{sec:experiment}
	To demonstrate that DiffuseMorph generates high-quality deformed images from the moving to the fixed images, we apply our method to the various image registration tasks. We conduct the experiments on the intra-subject image registration using 2D facial expression images and 3D cardiac MR scans. Also, we apply our model to 3D brain MR registration, in which individual brain images are deformed to a common atlas. The datasets and training details are as follows, and more details are described in Supplementary Material.

	\subsubsection{Datasets}
	For 2D face images, Radboud Faces Database (RaFD) \cite{langner2010presentation} was used. It contains 8 facial expressions collected from 67 subjects: neutral, angry, contemptuous, disgusted, fearful, happy, sad, and surprised. For each expression, 3 different gaze directions are provided. We cropped the data to $640\times 640$, resized them into $128\times 128$, and converted the RGB images to gray scale. We divided the data by 53, 7, and 7 subjects for training, validation, and test, respectively. 
	
	For 3D cardiac MR scans, we used ACDC dataset \cite{bernard2018deep} that provides 100 4D temporal cardiac MRI data from the diastolic to systolic phases and segmentation maps at both ends of the phases. We resampled all scans with a voxel spacing of $1.5\times 1.5\times 3.15mm$, cropped them to $128\times 128\times 32$, and normalized the intensity into [-1, 1]. We used 90 and 10 scans for training and test.
	
	Also, we used OASIS-3 dataset \cite{lamontagne2019oasis} for 3D brain MR registration. It provides brain MR images and corresponding volumetric segmentation maps from FreeSurfer \cite{fischl2012freesurfer}. We used 1156 T1-weighted scans that preprocessed by image resampling to $256\times 256\times 256$ grid with 1$mm^3$ isotropic voxels, affine spatial normalization, and brain extraction. The images were cropped by $160\times 192\times 224$. We used 1027, 93, and 129 scans for training, validation, and test, respectively.
	
	\subsubsection{Implementation Details}
	Our model was implemented using PyTorch library in Python. We used the network architecture designed in DDPM \cite{ho2020denoising} for the diffusion network, and set the noise level from $10^{-6}$ to $10^{-2}$ by linearly scheduling with $T_{train}=2000$. Also, we used the backbone of VoxelMorph-1 \cite{balakrishnan2018unsupervised} for the deformation network. Here, we configured layers of the networks according to the dimension of the image, e.g. 2D conv layer for 2D image registration. 
	For the face dataset, we set the hyper-parameter as $\lambda=2$, and trained the model with the learning rate $5\times 10^{-6}$ for 40 epochs. For the cardiac MR data, we trained the model with $\lambda=20$ and the learning rate $2\times 10^{-4}$ for 800 epochs. Also, we trained the model using the brain MR data for 60 epochs with $\lambda=10$ and the learning rate $1\times 10^{-4}$. Using a single Nvidia Quadro RTX 6000 GPU, we trained our model by Adam optimization algorithm \cite{kingma2015adam}.
	
	\subsubsection{Evaluation}
	To evaluate the registration performance, we computed the percentage of non-positive values of Jacobian determinant on the registration field ($|J_\phi|\!\leq\! 0$), which indicates that one-to-one mapping of the registration has been lost. Here, for the facial images, we measured NMSE and SSIM between deformed and fixed images. For the MR images, we computed Dice score between the deformed segmentation maps and the ground-truth labels for several anatomical structures. On the other hand, to evaluate the continuous deformation quality of cardiac scans, we computed PSNR and NMSE between the deformed images and the real data. For the comparative learning-based models, we used the same deformation network architecture and parameters for a fair comparison.

	\begin{figure*}[t!]
		\centering
		\includegraphics[width=\linewidth]{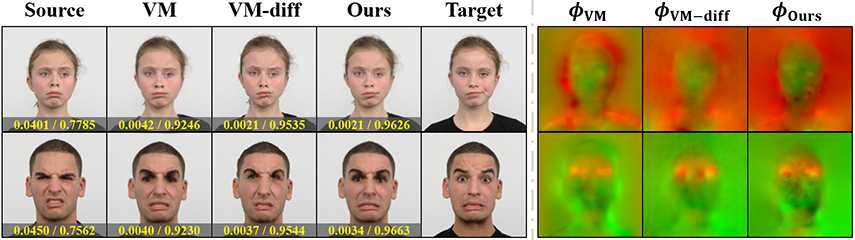}
		\caption{Visual comparison results of image registration on the facial images (left) using the estimated registration fields (right). Results are deformed from the right-gazed sad to the front-gazed contemptuous images (top), and from the left-gazed disgusted to the front-gazed fearful images (bottom). The average values of NMSE / SSIM are displayed on each registration result.}
		\label{fig:result_face}
	\end{figure*}

	\begin{table}[t!]
		\centering
		\caption{Quantitative evaluation results of the facial expression image registration. Standard deviations are shown in parentheses.}
		\resizebox{0.85\linewidth}{!}{
			\begin{tabular}{M{2.2cm}M{2.7cm}M{2.7cm}M{2.7cm}}
				\toprule
				Method & NMSE $\times 10^{-1}$ & SSIM & $|J_\phi|\leq 0$ (\%)\\
				\midrule
				\multicolumn{1}{l}{Initial} & 0.363 (0.268) & 0.823 (0.066) & 0 \\
				\multicolumn{1}{l}{VM \cite{balakrishnan2018unsupervised}} & 0.047 (0.057) & 0.936 (0.024) & 0.050 (0.106) \\
				\multicolumn{1}{l}{VM-diff \cite{dalca2018unsupervised}} & 0.034 (0.015) & 0.957 (0.013) & \textbf{0.014 (0.065)}  \\
				\multicolumn{1}{l}{Ours} & \textbf{0.032 (0.017)} & \textbf{0.964 (0.011)} & 0.017 (0.056) \\
				\bottomrule
		\end{tabular}  }
		\label{tab:result_face}
	\end{table}

	\subsection{Results of Intra-subject 2D Face Image Registration}
	We compared DiffuseMorph against VM \cite{balakrishnan2018unsupervised} and VM-diff \cite{dalca2018unsupervised}. We tested the image registration performance on the deformed images in RGB scale by applying the registration field to each RGB channel. Fig.~\ref{fig:result_face} shows visual comparisons of the registration results. Compared to VM and VM-diff, our model deforms the source image to be more accurately aligned with the target image. Also, as reported in Table~\ref{tab:result_face}, our model achieves lower NMSE and higher SSIM. Moreover, the metric of Jacobian determinant on registration fields of ours shows comparable values to VM-diff with the diffeomorphic constraint. These results indicate that the proposed DiffuseMorph provides high-quality image registration. More results of facial expression images can be found in Supplementary Material. 
	
\begin{figure}[t!]
		\centering
		\includegraphics[width=\linewidth]{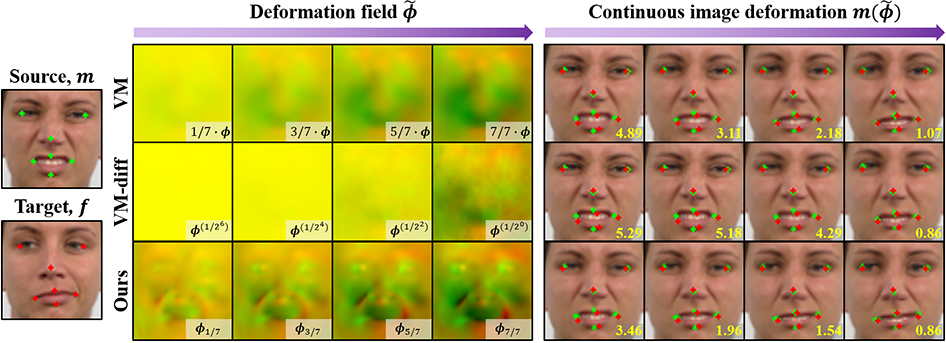}
		\caption{Results of continuous image deformation of the facial images with facial landmarks. The deformation is performed from the right-gazed disgusted to the left-gazed contemptuous images. The average of MSE between the deformed and target landmarks is displayed on each result.}
		\label{fig:result_face_int}
	\end{figure}

	\paragraph{Continuous image deformation}
	We also performed continuous deformations of the facial expression from the moving source to the fixed target. Fig.~\ref{fig:result_face_int} shows the intermediate images of our model and the comparative methods. We obtained the results of VM by scaling the registration field, i.e. $\zeta \cdot \phi$ with $0\leq \zeta \leq 1$, and those of VM-diff by integrating the velocity field along timescales, i.e. $\phi^{1/{2^v}}$ where $v$ is the number of time steps. We can see that the estimated registration fields of VM only vary in their scale, but the relative spatial distribution does not change. Also, VM-diff does not provide regularly continuous deformation. On the other hand, in the proposed method, the registration field changes non-uniformly according to $\eta$, depending on the importance of variations at the intermediate deformation level. The improved performance of our method also can be also quantitatively verified using the facial landmarks extracted by \textit{dlib} library in Python. Specifically, the visual results and MSE values in Fig.~\ref{fig:result_face_int} show that our model provides superior performance for the continuous deformation, which indicates the importance of the conditional score for deformation.

	\paragraph{Synthetic deformed image generation}
	\begin{wrapfigure}[17]{r}{0.45\textwidth}
		\begin{minipage}[t][5.5cm][c]{0.45\textwidth}
			\centering
			\includegraphics[width=\linewidth]{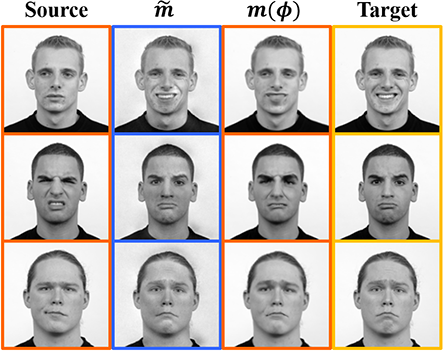}
			\caption{Results of the synthetic deformed image generation via our generative process. The process of the samples is included in Supplementary Material.}
			\label{fig:result_face_gen}
		\end{minipage}
	\end{wrapfigure}
	To verify the capability of image generation from our DiffuseMorph, we evaluated the generation using the facial images. Here, the generative process was performed using the unseen images in the training phase. Fig.\ref{fig:result_face_gen} shows the generated samples $\tilde{m}$ given the moving source and fixed target images. The samples are obtained from the noisy moving image with the noise level $\alpha_{200}$ for the forward diffusion. We set the number of reverse diffusion steps to $80$. As shown in the results, our model provides the synthetic deformed images similar to the target images for various pairs of facial expressions. Also, when compared to the warped image $m(\phi)$ using the registration field, we can observe that the proposed generative process is effective to provide image deformation if the moving image does not have teeth shown in the fixed image.
	
	\begin{table}[b!]
		\caption{Quantitative comparison results of the cardiac image registration. Standard deviations are shown in parentheses.}
		\label{tab:result_cardiac}
		\centering
		\resizebox{\linewidth}{!}{
			\begin{tabular}{M{1.5cm}M{2.4cm}M{2.3cm}M{2.4cm}M{2.35cm}M{1.55cm}}
				\toprule
				\multirow{2}{*}{Method} & \multicolumn{2}{c}{Image registration} & \multicolumn{3}{c}{Continuous deformation} \\ \cmidrule(l){2-3} \cmidrule(l){4-6}
				& Dice & $|J_\phi|\leq 0$ (\%) & PSNR (dB) & NMSE $\times 10^{-8}$ & Time (sec)  \\
				\midrule
				\multicolumn{1}{l}{Initial} & 0.642 (0.188) & - & 28.058 (2.205) & 0.790 (0.516) & - \\
				\multicolumn{1}{l}{VM \cite{balakrishnan2018unsupervised}}      & 0.787 (0.113) & 0.169 (0.109) & 30.678 (2.652) & 0.477 (0.453) & 0.219 \\
				\multicolumn{1}{l}{VM-diff \cite{dalca2018unsupervised}} & 0.794 (0.104) & 0.291 (0.188) & 29.481 (2.473) & 0.602 (0.477) & 2.902 \\
				\multicolumn{1}{l}{Ours} & \textbf{0.802 (0.109)} & \textbf{0.161 (0.082)} & \textbf{30.725 (2.579)} & \textbf{0.466 (0.432)} & 0.456 \\
				\bottomrule
		\end{tabular}}
	\end{table}
	
	\begin{figure}[b!]
		\centering
		\includegraphics[width=0.9\linewidth]{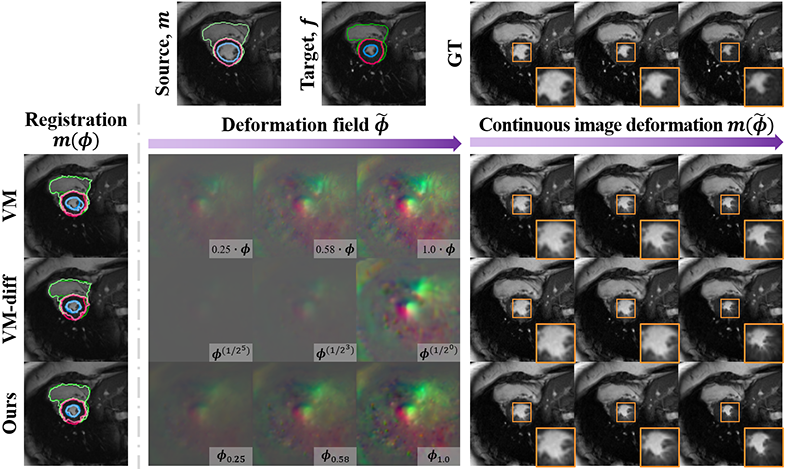}
		\caption{Results of the image registration and continuous deformation on the cardiac MR images. The registration results show the overlaid contours of segmentation maps (green: RV, red: Myo, blue: BP). GT is the ground-truth data.}
		\label{fig:result_cardiac}
	\end{figure}
	
	\subsection{Results of Intra-subject 3D Cardiac MR Image Registration}
	We tested the image registration of the end-diastolic images aligned with the end-systolic images. Table~\ref{tab:result_cardiac} reports the registration results with the average Dice score for the segmentation maps of left blood pool (BP), myocardium (Myo), left ventricle (LV), right ventricle (RV), and these total region, as well as the Jacobian metrics. Compared to the baseline methods of VM \cite{balakrishnan2018unsupervised} and VM-diff \cite{dalca2018unsupervised}, our model achieves high registration accuracy with a comparable number of folds in topology preservation.
	
	In addition, since the cardiac dataset we used provides 4D data between the end-diastolic to the end-systolic phases, we performed quantitative evaluation for the continuous deformation. As shown in Table~\ref{tab:result_cardiac}, when we measured the PSNR and NMSE between the deformed images and ground-truth reference images, our method provides continuous deformed images more similar to the ground-truth images than the comparative methods.
	The visual comparison results in Fig.~\ref{fig:result_cardiac} also shows the superiority of our method. More visual results can be found in Supplementary Material.
	
	\begin{figure}[b!]
		\centering
		\includegraphics[width=\linewidth]{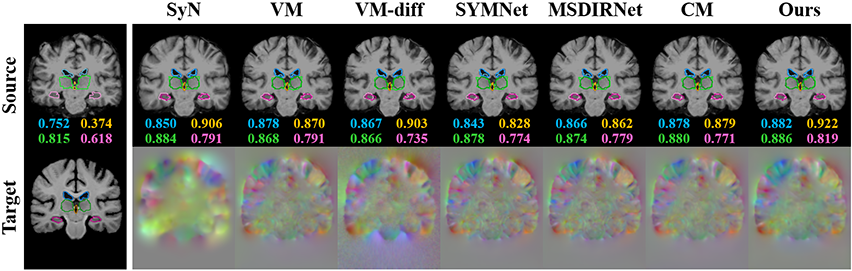}
		\caption{Results of the image registration on the brain MR images (top) and the estimated registration fields (bottom). Segmentation maps of several anatomical structures are overlaid with the contours (blue: ventricles, green: thalami, orange: third ventricle, pink: hippocampi). The Dice scores for each structure are displayed with the corresponding colors on each result.}
		\label{fig:result_brain}
	\end{figure}

	\begin{figure}[b!]
		\centering
		\includegraphics[width=\linewidth]{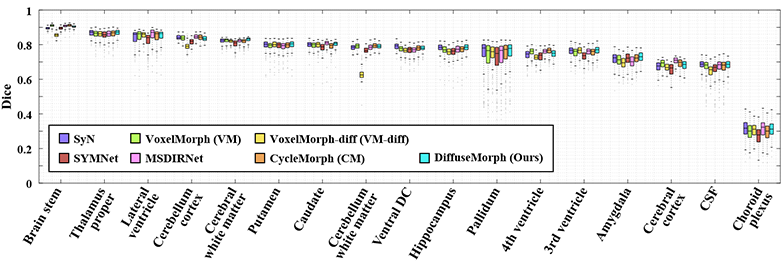}
		\caption{Quantitative evaluation results of Dice scores on brain anatomical structures in the brain MR image registration experiment.}
		\label{fig:result_brain_quan}
	\end{figure}
	
	\subsection{Results of Atlas-based 3D Brain MR Image Registration}
	For the brain MR image registration, we compared our DiffuseMorph to the following comparative methods: SyN \cite{avants2008symmetric}, VM \cite{balakrishnan2018unsupervised}, VM-diff \cite{dalca2018unsupervised}, SYMNet \cite{mok2020fast}, MSDIRNet \cite{lei20204d}, and CM \cite{kim2021cyclemorph}. As shown in the visual comparison results of Fig.\ref{fig:result_brain}, our model estimates smooth registration fields and yields deformed moving images that are more accurately aligned to the fixed images, compared to the baseline models. This also can be observed through the contours of the segmentation maps of several brain anatomical structures. We provided more visual results in Supplementary Material. Fig.~\ref{fig:result_brain_quan} and Table~\ref{tab:result_brain} report the results of quantitative evaluation. These show that the proposed method achieves higher Dice scores with less non-positive values of the Jacobian determinant over the existing learning-based methods, which empirically suggests that the proposed method can provide accurate image registration with improved topology preservation.

	\begin{table}[t!]
		\centering
		\caption{Quantitative evaluation results of the brain MR image registration. Standard deviations are shown in parentheses.}
		\resizebox{0.85\linewidth}{!}{
			\begin{tabular}{M{2.3cm}M{2.7cm}M{2.7cm}M{2.6cm}}
				\toprule
				Method & Dice & $|J_\phi|\leq 0$ (\%) & Time (min) \\
				\midrule
				\multicolumn{1}{l}{Initial} & 0.616 (0.171) & 0  & 0 \\
				\multicolumn{1}{l}{SyN \cite{avants2008symmetric}} & 0.752 (0.140) & 0.400 (0.100) & 122, CPU \\
				\multicolumn{1}{l}{VM \cite{balakrishnan2018unsupervised}} & 0.749 (0.145) & 0.553 (0.075) & 0.01, GPU \\
				\multicolumn{1}{l}{VM-diff \cite{dalca2018unsupervised}} & 0.731 (0.139) & 0.631 (0.073) & 0.01, GPU \\
				\multicolumn{1}{l}{SYMNet \cite{mok2020fast}} & 0.733 (0.148) & 0.547 (0.049) & 0.43, GPU  \\
				\multicolumn{1}{l}{MSDIRNet \cite{lei20204d}} & 0.751 (0.142) & 0.804 (0.089) & 2.06, GPU  \\
				\multicolumn{1}{l}{CM\cite{kim2021cyclemorph}} & 0.750 (0.144) & 0.510 (0.087) & 0.01, GPU \\
				\multicolumn{1}{l}{Ours} & \textbf{0.756 (0.139)} & \textbf{0.505 (0.058)} & 0.01, GPU \\
				\bottomrule
		\end{tabular}}
		\label{tab:result_brain}
	\end{table}

	\subsubsection{Study on the Effect of Registration Loss} 
	\begin{wraptable}[9]{r}{0.5\linewidth}
		\begin{minipage}[t][2cm][c]{0.5\textwidth}
			\centering
			\caption{Result of the study on the effect of registration loss. Standard deviations are shown in parentheses.}
			\resizebox{\linewidth}{!}{
				\begin{tabular}{M{1.4cm}M{2.6cm}M{2.6cm}}
					\toprule
					Method & Dice & $|J_\phi|\leq 0$ (\%) \\
					\midrule
					\multicolumn{1}{l}{$\lambda=2$} & 0.736 (0.152) & 0.498 (0.098) \\
					\multicolumn{1}{l}{$\lambda=5$} & 0.746 (0.143) & 0.499 (0.070) \\
					\multicolumn{1}{l}{$\lambda=10$} & 0.756 (0.139) & 0.505 (0.058)  \\
					\bottomrule
			\end{tabular}  }
			\label{tab:result_lambda}
		\end{minipage}
	\end{wraptable}
	To explore the effect of the registration loss on the performance of our model, we performed a comparison study by varying the value of $\lambda$ in (\ref{eq:loss_tot}) using brain MR data. As reported in Table~\ref{tab:result_lambda}, when $\lambda$ is lower, the Dice score decreased but  produced better regularity of registration fields.  This indicates that the registration loss forces our model to provide more accurate image registration, but the trade-off needs to be balanced.

	\section{Conclusion}
	\label{sec:Conclusion}
	We presented a novel DiffuseMorph model for unsupervised image registration by employing the diffusion probabilistic model that is jointly trained with the deformation network. Thanks to the capability of learning the conditional score function of deformation, the proposed method not only generates synthetic deformed images but also provides high-quality image registration from the continuous deformation by estimating registration fields along the trajectory for the moving image toward the fixed image. We expect that DiffuseMorph can be a promising algorithm to generate temporal data using moving and fixed images.
	
	\subsubsection{Acknowledgments}
	This work was supported in part by the National Research Foundation (NRF) of Korea under Grant NRF-2020R1A2B5B03001980, in part by Field-oriented Technology Development Project for Customs Administration through NRF of Korea funded by the Ministry of Science \& ICT and Korea Customs Service(NRF-2021M3I1A1097938), and in part by the KAIST Key Research Institute (Interdisciplinary Research Group) Project. 
	%

\clearpage
\appendix
\section{Details of Implementation}
\label{sec:implementation}

\subsection{Network Architecture}
\subsubsection{Diffusion Network}
\begin{figure}[b!]
	\centering
	\includegraphics[width=0.96\linewidth]{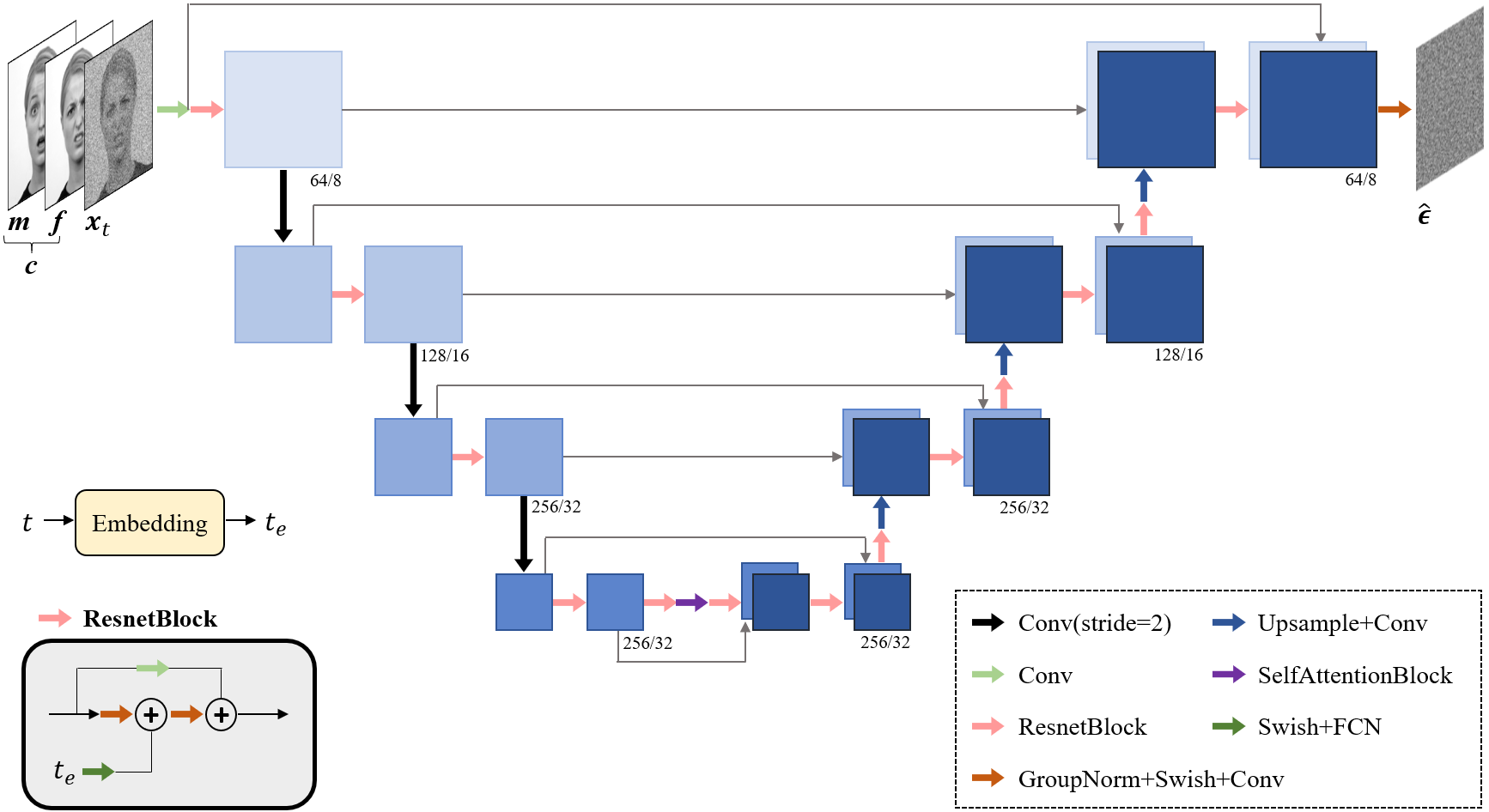}
	\caption{The architecture of the diffusion network $G_\theta$. The Resnet block (pink arrow) is composed of the group normalization, swish function, and convolution layers. The number of convolutional channels is shown as $a/b$ below the square box, where $a$ and $b$ are for 2D and 3D networks, respectively. The downsampling in the encoder is performed by the convolution layer with stride 2, whereas the upsampling in the decoder is done by nearest interpolation with scale factor 2, followed by the convolution layer with stride 1. Each feature map of the encoder is given to the decoder via skip-connection.  }
	\label{fig:sup_network_diffuse}
\end{figure}

As we described in the paper, we employed the network architecture of DDPM \cite{ho2020denoising} for the diffusion network $G_\theta$ of the proposed method. Fig.~\ref{fig:sup_network_diffuse} illustrates the structure of the diffusion network. Specifically, it consists of four encoder and four decoder blocks. Each block has the Resnet block composed of the group normalization \cite{wu2018group}, the swish function \cite{elfwing2018sigmoid}, and convolution layers, which takes the embedded time $t_e$ as well as the output of the preceding layer. At the last encoder block, the feature maps are attended by the self-attention module \cite{vaswani2017attention}. Also, each decoder block additionally takes the feature maps of the encoder outputs via skip-connection that enables the decoder to use the encoding information of inputs. Accordingly, when the moving image $m$, fixed image $f$, and perturbed target image $x_t$ are given to the diffusion network with the time step $t$, the network is trained to estimate the latent feature of the conditional score function $\hat{\epsilon}$ of the deformation between the moving and fixed images. Here, we configured the kernel dimension of convolution layers according to the input image dimension. 

\subsubsection{Deformation Network}
\begin{figure}[h!]
	\centering
	\includegraphics[width=0.96\linewidth]{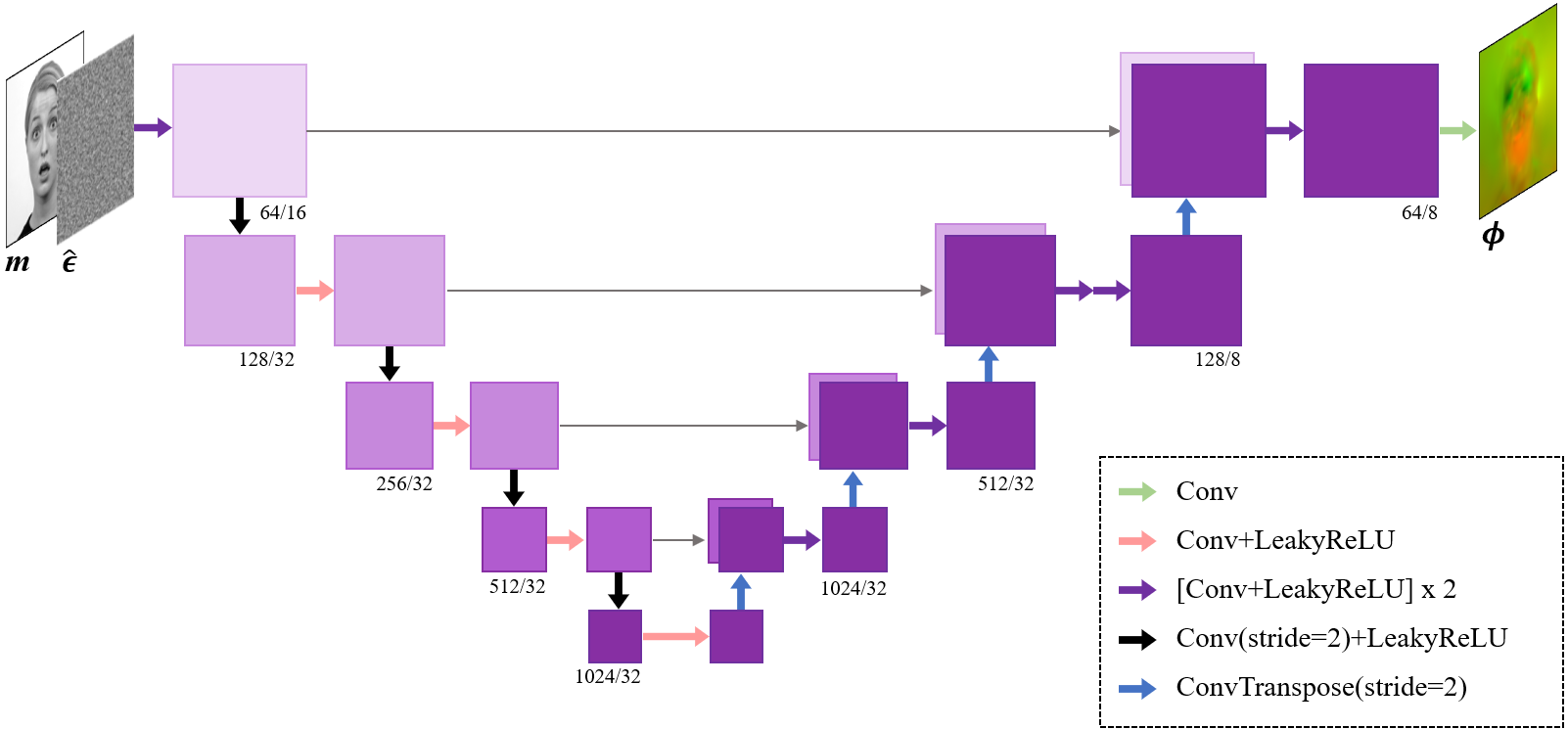}
	\caption{The architecture of the deformation network $M_\psi$. It is a modified U-net structure with the convolution layer and leakyReLU activation function. The number of convolutional channels is shown as $a/b$ below the square box, where $a$ and $b$ are for 2D and 3D networks, respectively. The image features are downsampled by the convolution with stride 2 (black arrow), while they are upsampled by the transposed convolution with stride 2 (blue arrow). The feature map of each encoder block is concatenated to that of the decoder block at the same level. }
	\label{fig:sup_network_deform}
\end{figure}

For the deformation network $M_\psi$, we implemented VoxelMorph-1 \cite{balakrishnan2018unsupervised} that presents the network architecture for image registration. As illustrated in Fig.~\ref{fig:sup_network_deform}, the deformation network is a U-shape network with encoder and decoder blocks, similar to the diffusion network. However, instead of the Resnet block, each block has the convolution and leakyReLU \cite{he2015delving} layers, called CL units, which enables the network to focus on image processing and learn the complex image features. Also, as the downsampling by the convolution is with stride 2, the upsampling is performed by the transposed convolution with stride 2. The final network output is generated by the convolution with stride 1. Thus, given the moving image $m$ and the latent feature of the diffusion network output $\hat{\epsilon}$, the deformation network estimates the registration field $\phi$ that warps the moving image into the fixed image. Similar to the diffusion network, the convolution layers are configured depending on the input image dimension.

\subsection{Data processing}
The intensity range of grayscale facial data and medical images used in the experiments of the main paper is [0, 1]. We scaled this intensity range of the data into [-1, 1]. Then, the noisy target is sampled using the scaled image and given to our model as an input, along with a condition of the moving and fixed images. Since the moving image is deformed by the registration field using the spatial transformation layer with linear interpolation, we rescaled the moving image into [0, 1] just before warping the moving image. For data augmentation of the facial data, we used random horizontal flipping. On the other hand, for the brain and cardiac MR data, we used random horizontal/vertical flips and random rotations with 90 degrees for data augmentation.

\begin{figure}[b!]
	\centering
	\includegraphics[width=\linewidth]{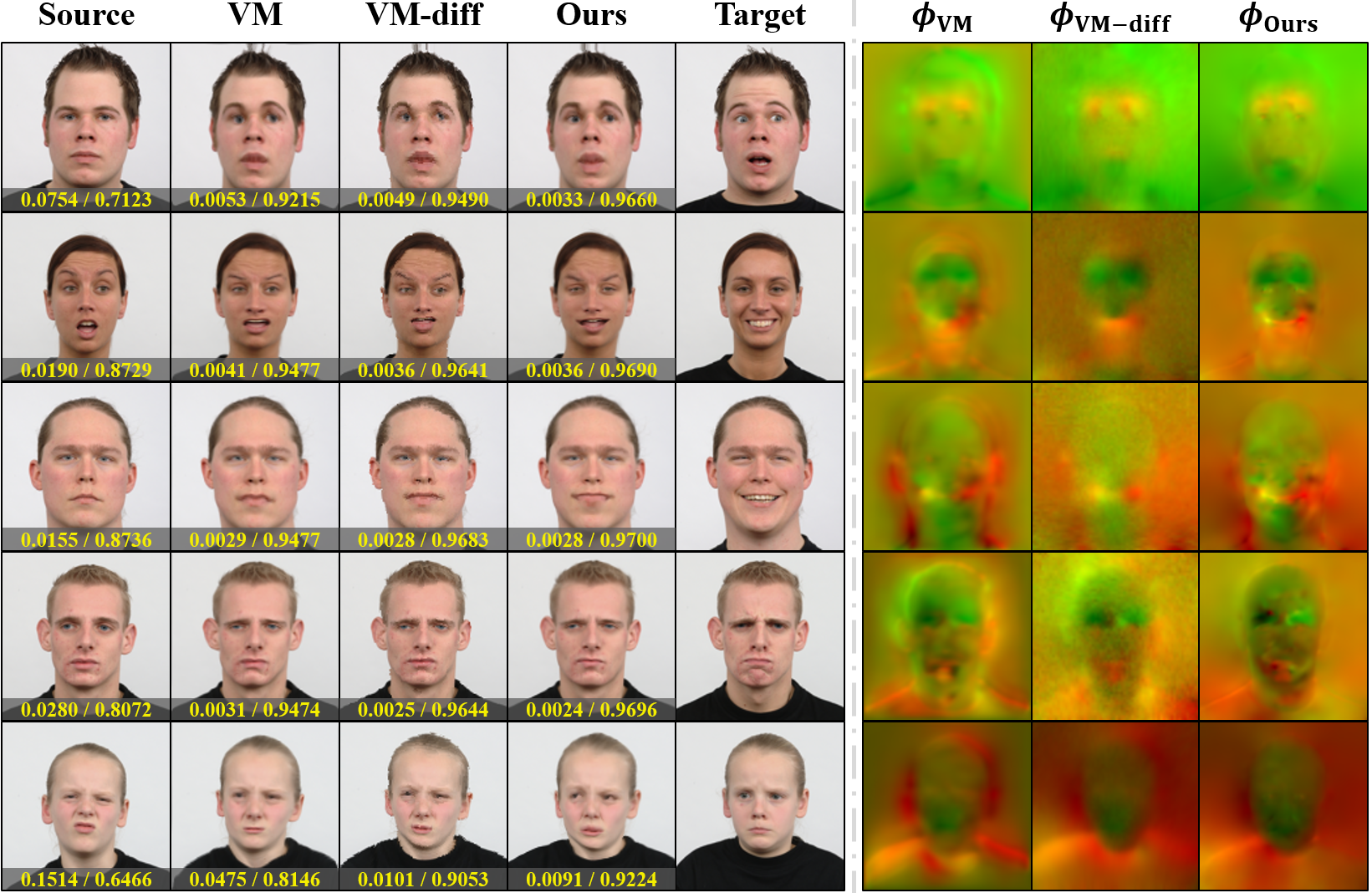}
	\caption{Visual results of facial expression image registration (left) using the estimated registration fields (right). From top to bottom, the results are deformed from the front-gazed neutral to right-gazed surprised images, from the right-gazed surprised to front-gazed happy images, from the front-gazed neutral to right-gazed happy images, from the left-gazed neutral to front-gazed angry images, from the front-gazed disgusted to left-gazed sad images. The mean values of NMSE/SSIM are displayed on each result.} 
	\label{fig:sup_face_com}
\end{figure}

\section{Additional experimental results}
\label{sec:experiment}

\subsection{Image Registration Results of the Comparisons}
\subsubsection{2D Facial Expression Image Registration}
For the intra-subject facial expression image registration task, we compared the proposed method with the VoxelMorph (VM) \cite{balakrishnan2018unsupervised} and VM-diff \cite{dalca2018unsupervised}. We implemented these methods using the same architecture of the deformation network $D_\psi$ of our model and trained the networks until the training loss converges for a fair comparison. Fig.~\ref{fig:sup_face_com} shows the results of visual comparisons on various facial expression images. Compared to the other baseline methods, we can see that our model provides more accurate deformation of the moving source image into the fixed target image by the smooth registration field. This can be also observed through the quantitative evaluation with NMSE and SSIM that are displayed on each result. 

\begin{figure}[b!]
	\centering
	\includegraphics[width=\linewidth]{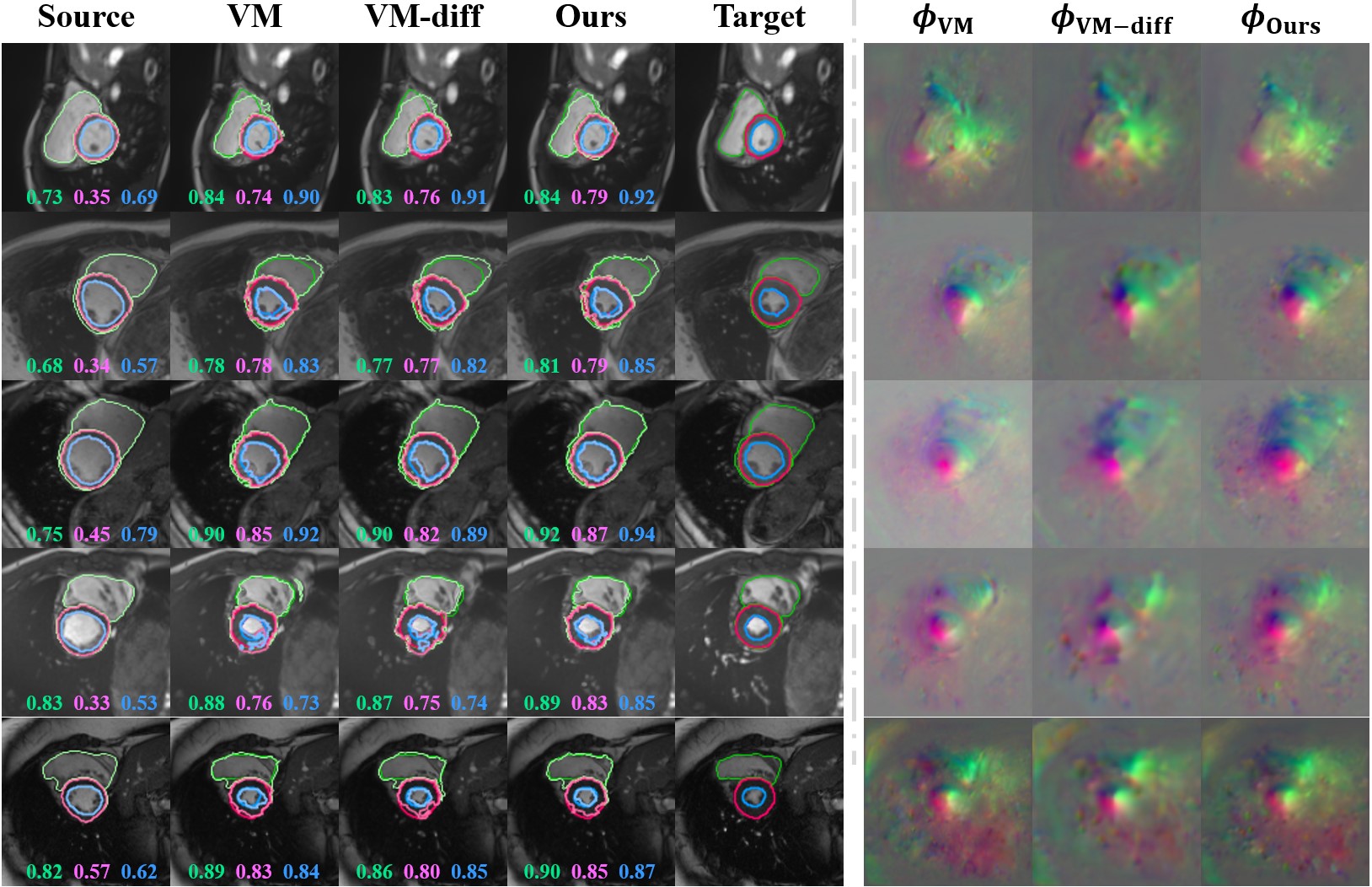}
	\caption{Visual results of cardiac MR image registration (left) using the estimated registration fields (right). The registration results show the overlaid contours of segmentation maps (green: epicardium of the right ventricle (RV), red: myocardium of left ventricle (Myo), blue: left blood-pool (BP)). The Dice score for each structure is displayed with the corresponding color on each result.} 
	\label{fig:sup_cardiac_com}
\end{figure}

\subsubsection{3D Cardiac MR Image Registration}
We also implemented our method for the intra-subject 3D cardiac MR image registration task. For the baseline methods, we compared ours with VM \cite{balakrishnan2018unsupervised} and VM-diff \cite{dalca2018unsupervised}. As the face image registration, we trained these models using the deformation network architecture $D_\psi$ of our method. Fig.~\ref{fig:sup_cardiac_com} visualizes the registration results of the cardiac image at the end-diastolic phase to the end-systolic phase. We also display the contours of segmentation maps for several structures and their Dice scores. The results show that our proposed method achieves higher registration performance than the comparative methods in that the moving source image is more accurately aligned with the fixed target image. 

\begin{figure}[b!]
	\centering
	\includegraphics[width=\linewidth]{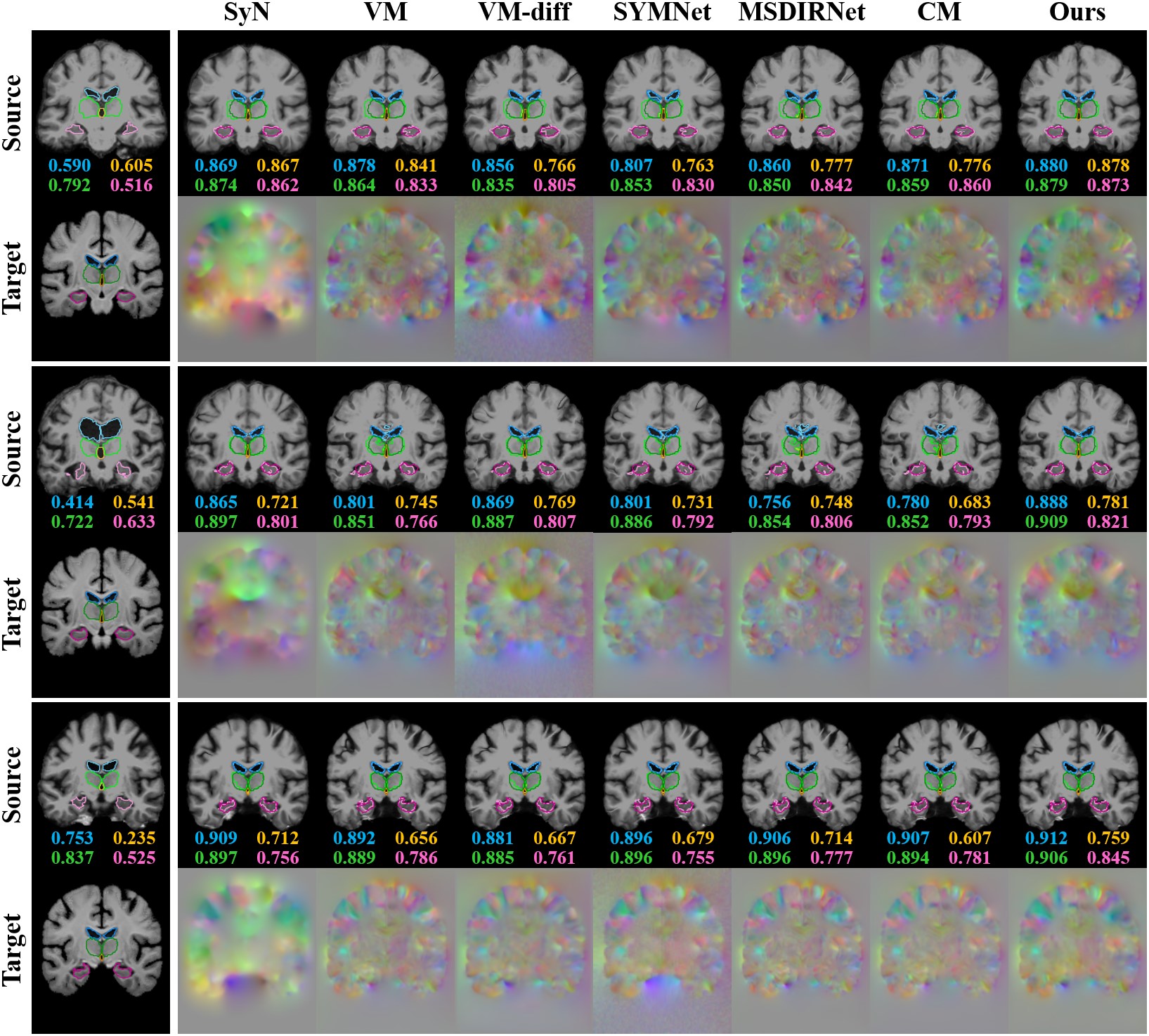}
	\caption{Visual results of atlas-based brain MR image registration (odd rows) using the estimated registration fields (even rows). Segmentation maps of several anatomical structures are overlaid with the contours (blue: ventricles, green: thalami, orange: third ventricle, pink: hippocampi). The Dice score of each structure is displayed with the corresponding color on each result.}
	\label{fig:sup_brain_com}
\end{figure}

\subsubsection{3D Brain MR Image Registration}
To verify the performance of the atlas-based 3D brain image registration, we employed the following comparative methods: SyN \cite{avants2008symmetric} by Advanced Normalization Tools (ANTs) \cite{avants2011reproducible}, VM \cite{balakrishnan2018unsupervised}, VM-diff \cite{dalca2018unsupervised}, SYMNet \cite{mok2020fast}, MSDIRNet \cite{lei20204d}, and CM \cite{kim2021cyclemorph}. For the learning-based methods, we used the 3D model of deformation network $D_\psi$ as a baseline network and set the same parameters for a fair comparison. Fig.~\ref{fig:sup_brain_com} shows the results of brain image registration. The Dice scores for several anatomical structures are displayed on each result, and the overall quantitative evaluation results can be found in the main paper. The visual results with the contours of segmentation maps show that the proposed DiffuseMorph deforms the moving source image more similar to the fixed target images than the others, not only in the overall shape but also in the detailed structures.

\begin{figure}[b!]
	\centering
	\includegraphics[width=\linewidth]{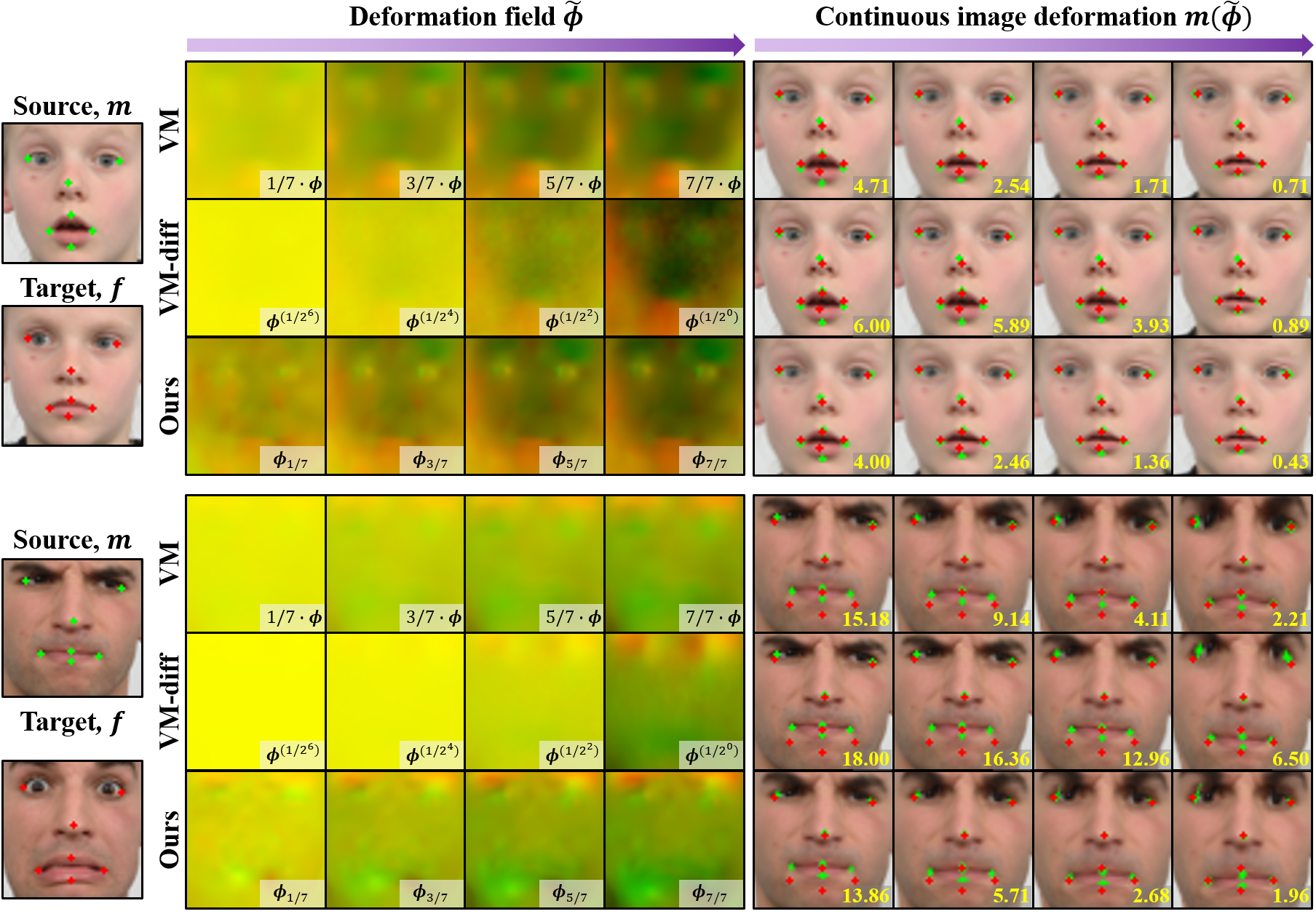}
	\caption{Results of the continuous image deformation using the facial expression images with landmarks. Image registration is performed from the front-gazed surprised to the left-gazed neutral images (top), and from the right-gazed angry to the front-gazed fearful images (bottom). The average of MSE between the deformed and target facial landmarks is displayed on each result. } 
	\label{fig:sup_face_int}
\end{figure}


\subsection{Image Registration Along Continuous Trajectory}
The proposed method can provide continuous image deformation for the moving image along the trajectory toward the fixed image, which is one of the main contributions of our paper. Here, we show additional results of ours and comparative methods, VM and VM-diff. For the case of VM, we obtain the deformations by linearly interpolating the registration field. For the VM-diff, the continuous deformation is done by integrating the velocity field in shorter timescales. In contrast, our method yields the intermediate images by scaling the latent feature from the conditional score function of the deformation. 

\begin{figure}[b!]
	\centering
	\includegraphics[width=\linewidth]{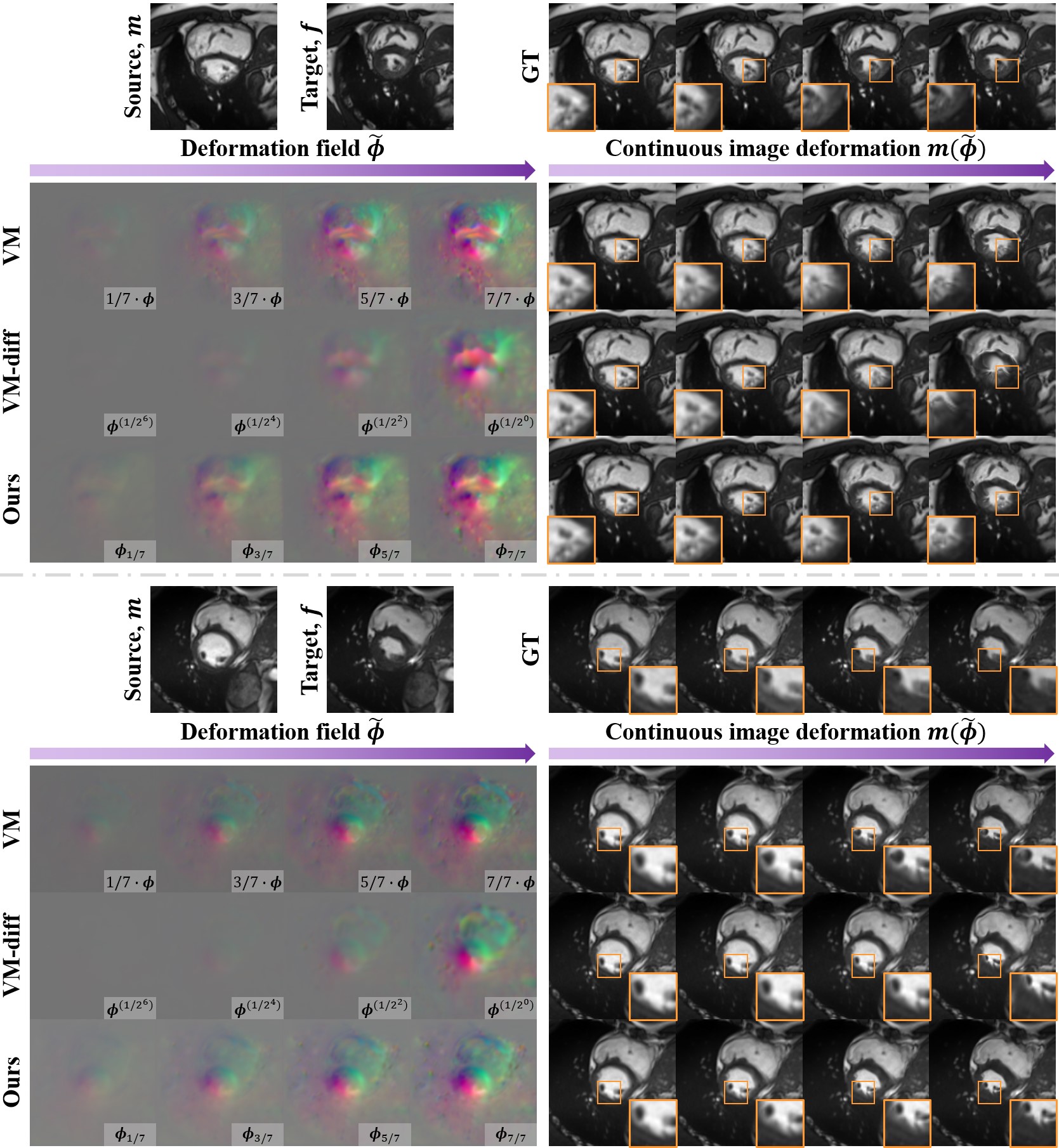}
	\caption{Results of the continuous image deformation using the cardiac MR images. GT is the ground-truth data, and the orange box shows the remarkable regions.} 
	\label{fig:sup_cardiac_int}
\end{figure}

Fig.~\ref{fig:sup_face_int} shows the results on the facial expression image registration task. As can be seen from the results, the estimated registration fields of VM are only scaled so that the change of the specific facial movement such as eyes is not clearly visible. Also, VM-diff is limited in providing continuous deformation since the registration fields in early levels are near zero but deform images rapidly in late levels. On the other hand, in our proposed method, the estimated registration fields from the scaled latent feature are not just a scaled version as in the VM, but rather exhibits very dynamically changing deformation fields depending on the positions (for example, more specifically consistent movement along the eyes and mouths compared to other methods). Thus, the resulting intermediate deformed images of our method have distinct changes from the moving and fixed images. These results can be observed similarly in Fig.~\ref{fig:sup_cardiac_int}, which visualizes the comparison results for the cardiac MR image registration task and verifies the continuous deformation performance of our method.

\begin{figure}[b!]
	\centering
	\includegraphics[width=0.95\linewidth]{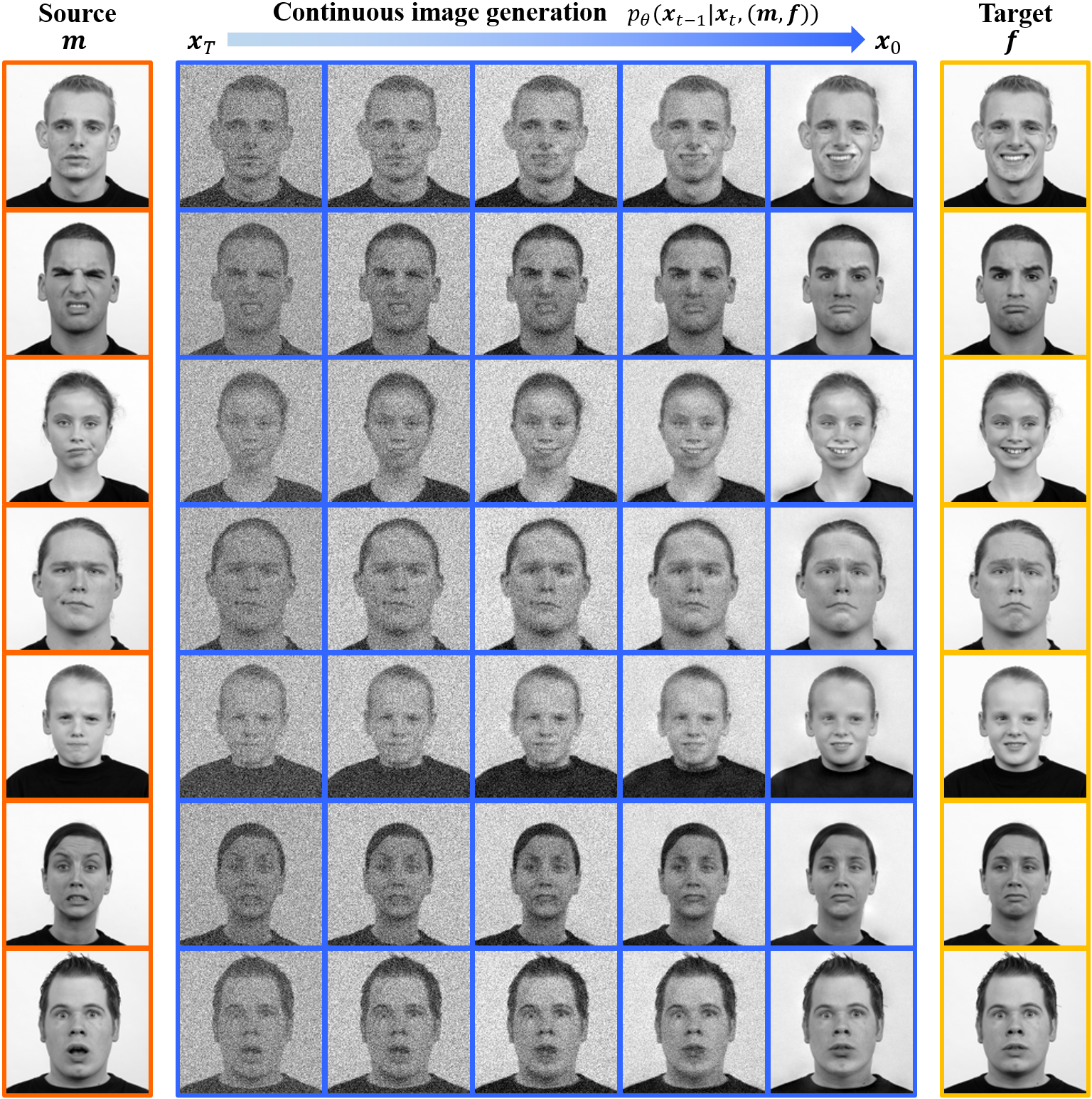}
	\caption{Results of the synthetic deformed image generation via our generative process from $T=80$. From top to bottom, the deformed image is generated from the left-gazed neutral to the front-gazed happy images, from the front-gazed disgusted to the front-gazed sad images, from the front-gazed contemptuous to the right-gazed happy images, from the front-gazed contemptuous to the front-gazed sad images, from the front-gazed angry to the right-gazed happy images, from the front-gazed fearful to the left-gazed sad images, from the front-gazed surprised to the front-gazed fearful images.}
	\label{fig:sup_face_gen}
\end{figure}

\subsection{Synthetic Deformed Image Generation}
In addition to the image registration, thanks to jointly training of the diffusion and deformation networks, our DiffuseMorph provides the image generation via the reverse of the diffusion process. As described in the main paper, given the condition with a pair of moving and fixed images, the generation process starts from one step forward diffusion on the moving image. Then, the noisy moving image with a certain noise level is refined iteratively by the reverse diffusion steps, resulting in the synthetic deformed images aligned with the fixed image. Fig.~\ref{fig:sup_face_gen} shows the generative process on the facial expression images. The sampling results are obtained by 80 diffusion steps starting from the moving image with the noise level $\alpha_{200}$. As our method learns the conditional score function of the deformation using various pairs of facial expression images, we can see that the generated samples from the moving images become similar to the fixed images. This indicates that our model has a capacity for conditional image generation as well as image registration.

\end{document}